# Single-shot Fresnel incoherent digital holography based on geometric phase lens

DONG LIANG,[1,2,3] QIU ZHANG,[2,3] AND JUN LIU[2,*]

[1]School of Physics Science and Engineering, Tongji University, Shanghai 200092, China
[2]Shanghai Institute of Optics and Fine Mechanics, Chinese Academy of Sciences, Shanghai 201800, China
[3]University of Chinese Academy of Sciences, Beijing 100049, China
*Corresponding author: jliu@siom.ac.cn

Single-shot and incoherent imaging are two important properties for digital holography technique to extend its application range. In this paper, a simple and compact in-line Fresnel incoherent digital holographic recording system with capability of single-shot and three-dimensional imaging is proposed. A GP lens is used as the common-path interferometer based on its special wave separation property. Parallel phase-shifting holography is used to achieve single-shot operation by using a polarization imaging camera together with space-division multiplexing. The capability of the proposed technique and the setup was experimentally confirmed by images of two standard test charts.

Digital holography is a powerful technique capable of recording 3D information using an image sensor and retrieving the amplitude and phase of an object in a computer [1]. It has been widely used in lots of research fields: 3D optical display [2], object recognition [3], quantitative phase imaging of biological samples [4], and so on. Among digital holography, the incoherent digital holography is one of the important research fields because it can avoid the speckle noise and spurious interference owing to the use of incoherent light source. Furthermore, it greatly expands the application field of digital holography to imaging with incoherent light such as fluorescence microscopy and white light imaging [5, 6]. In incoherent holography, unlike multiple view projection holography [7], scanning holography [8], and other incoherent holography methods, Fresnel incoherent correlation holography (FINCH) [9] is a non-scanning and motionless method during obtaining holograms, and it got more and more attention recently. In FINCH, the spatial light modulator (SLM) splits the spherical wave originated from each object point into two spherical waves with different curvatures and recombined at an imaging plane to produce interference fringes. Therefore, separating wave is an important step in the FINCH system. The emergence of high-performance SLM provides a flexible and convenient method to separate wave for this incoherent holography. Owing to the use of coaxial holographic recording optical configuration, the phase-shifting interferometry is used to effectively eliminate undesired diffraction images and twin images from an object wave. Thereafter, the noiseless images are obtained through a numerical reconstruction algorithm by using a computer.

By using SLM, FINCH has achieved many remarkable results on suppression of DC term and conjugate image [10, 11], improvement of imaging resolution and the reconstructed image quality [12, 13] in the past decade. Until now, the phase-only SLM is the most common device used as a wave separator for a FINCH system owing to its convenient on the tunable phase pattern. However, this kind of FINCH system needs to capture at least three holograms to achieve one complex hologram free from twin image and zero-order disturbances. Additionally, the phase-only SLM increases the complexity and cost of the optical setup which will limit its broad range application. Recently, some FINCH systems based on the transmissive liquid crystal gradient index lens or birefringent crystal lens have also been proposed to simplify the setup [14, 15].

In this letter, the geometric phase (GP) lens is introduced to the FINCH system which can perfectly replace the above mentioned complex and expensive SLM or birefringent crystal lens [16, 17]. Together with a detector using a polarized camera, single-shot FINCH is achieved with very simple, compact and cost-effective setup.

The GP lens we used in FINCH system employs photo-aligned liquid crystal layers to implement the spatially varying Pancharatnam-Berry phase [18], leading to the expected polarization and focal length dependent wavelength. Owing to the propagation property of GP lens, left-hand circular polarization (LHCP) and right-hand circular polarization (RHCP) will focus with a fixed positive or negative focal length, respectively. As for linear polarized light and non-polarized light, the GP lens will also act as both a positive lens and a negative lens with a 50:50 ratio of the light. The principle of FINCH system with GP lens is shown in Figure 1. An ordinary convex lens is used to collect the linear polarized or non-polarized incoherent light from any point on the object plan and then pass through the GP lens. The GP lens divides the beam into a converging one and a diverging one simultaneously that will coincide on the image sensor of CCD to create an interferogram. The object image is obtained on the image plan through a mathmatical reconstruction algorithm. Moreover, the GP lens has high transmissive efficiency in a broadband spectral range with a simple structure and cheap price. Single lens imaging experiments among GP lens, plan-convex lens, Fresnel lens, and aspherized achromatic lens had showed that GP lens owned a high imaging quality [19]. Therefore, such a geometric-phase lens is a perfect wave separating optics for FINCH [9].

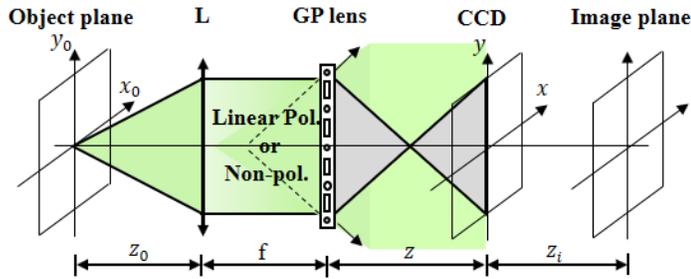

Fig. 1. Schematic diagram of the FINCH system. GP lens, geometric phase lens; CCD, charge-coupled device; L, convex lens; Linear Pol., linear polarized; Non-pol., non-polarized;

On the other hand, phase shifting is usually used in the FINCH by using phase-only SLM or liquid crystal variable retarder to avoid the bias and twin images issue. However, this design can not run in single-shot which is necessary in some fast and instantaneous measurements. Furthermore, the hologram collected in every step for phase-shifting method may disturbed by environment disturbance which will affect the quality of the reconstructed image. Parallel phase-shifting digital holography using a polarization imaging camera with a micro-polarizer array can capture a hologram in single-shot which contains the information of multiple phase-shifted interferograms [20, 21]. Figure 2 shows the principle of the parallel phase-shifting technique. Through the processes of demosaicing and interpolating of the single-shot recorded hologram, four-step phase shifting holograms are mathematically generated.

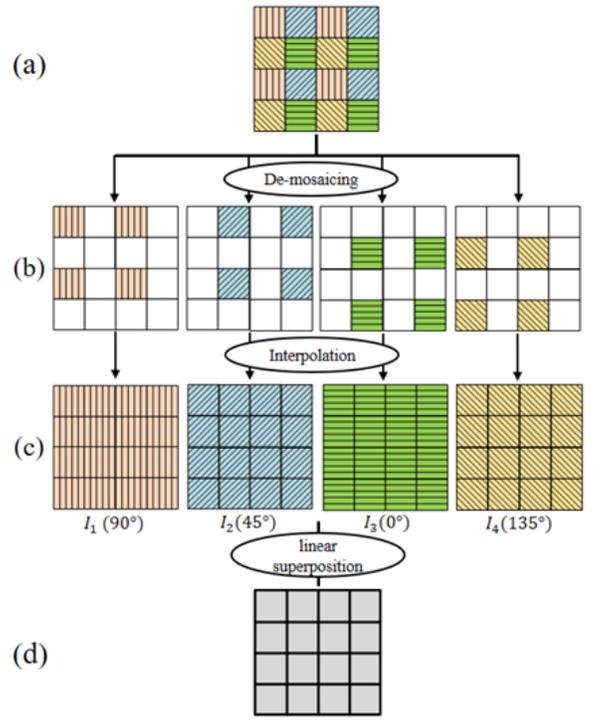

Fig. 2. Reconstruction process of single-shot phase-shifting holograms. (a) Recorded interference pattern image; (b) Extraction for each phase-shift; (c) Four holograms required for phase-shifting interferometry at 0°, 45°, 90° and 135° four phase-shifting angles; (d) Complex amplitude distribution of incoherent light.

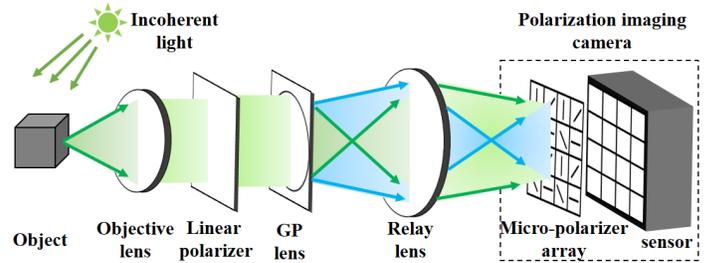

Fig. 3. Schematic diagram of the Single-shot FINCH system using GP lens.

An ordinary convex lens is used to collect the incoming incoherent light from objects and then pass through a linear polarizer and a GP lens arranged in order. After the GP lens, the beam is simultaneously divided into a converging one and a diverging one. Since the curvatures of the two beams modulated by the GP lens are relatively large, it is difficult to observe the interference pattern directly. To enhance the fringe pattern visibility, a relay lens is therefore employed after the GP lens. Finally, a polarization imaging camera is used to record the two-wave interference. The whole optical setup is in-line, simple and compact. Since the single-shot recorded interferogram including the information of four phase-shifted holograms with the phase shifts of 0°, 45°, 90°, and 135°, respectively, the complex hologram amplitude and phase data can be achieved by using typical four-step phase-shifting method [22], as shown by Eq. (1):

$$H(x,y) = \left(\frac{I_1-I_2}{1-e^{i\alpha}} - \frac{I_3-I_4}{e^{i\beta}-e^{i\theta}}\right) / \left[\frac{R^*(1-e^{-i\alpha})}{1-e^{i\alpha}} - \frac{R^*(e^{-i\beta}-e^{-i\theta})}{e^{i\beta}-e^{i\theta}}\right]$$

$$\Delta\emptyset(x,y) = \tan^{-1}\left(\frac{\sqrt{I_1+I_2-I_3-I_4}\cdot\sqrt{3I_2-3I_3-I_1+I_4}}{I_2+I_3-I_1-I_4}\right) \quad (1)$$

$I_1$, $I_2$, $I_3$ and $I_4$ in Eq. (1) denotes the mathematically generated four-step phase-shifted holograms with different phase values of 0, α, β, and θ, which are 0°, 45°, 90° and 135°, respectively. H(x, y) represents a complex amplitude distribution of incoherent light on the image sensor plane which is reconstructed by the linear superposition of four holograms. $\Delta\emptyset(x,y)$ is the phase of complex hologram. Then the original object image S(x, y) can be reconstructed from H(x.y) by calculating the Fresnel propagation formula as Eq. (2)

$$S(x,y) = H(x,y) * \exp\left[-\frac{i\pi}{\lambda z}(x^2+y^2)\right] \quad (2)$$

where * denotes a 2D convolution, z represents the reconstruction distance.

To experimentally prove our proposed single-shot FINCH, a proof-of-principle setup is built as shown in the photograph displayed in Figure 4.

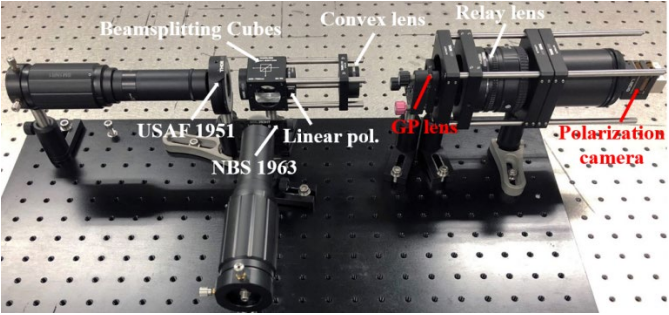

Fig. 4. The photograph of the experimental setup.

An United States Air Force (USAF) 1951 resolution test chart is used as the imaging target. The system is illuminated by a 530nm Fiber-Coupled LED (M530F2, Thorlabs) with a spectral bandwidth of about 30 nm. A convex lens with a focal length of 100 mm is used to collect the incoming incoherent light after pass through the USAF 1951 chart. A linear polarizer is located in front of the GP lens to introduce a linear polarization light for GP lens. The distance between the GP lens and the convex lens is about 80 mm. The GP lens utilized in the system herein is an off-the-shelf model (#34-466, Edmund optics) with a focal length of 100 mm. A polarized sensor (PHX050S-P, LUCID) is located about 120 mm behind the Nikon lens. The pixel number of the image sensor is 2448×2048 with a pixel pitch of 3.45 μm. It need to be noted that a micro-polarizer array is attached to the image sensor pixel by pixel. Four phase-shifted hologram patterns are recorded simultaneously in four parts of the polarization sensor in every single exposure. Furthermore, to investigate the 3D imaging capability of the setup, another 530nm Fiber-Coupled LED and an NBS 1963 chart with 4.5 cycle/mm region are setup normal to and then combined to the main optical path with a beam combiner, as shown in Figure 4. The distance between USAF 1951 and the beam combiner or NBS 1963 are about 30 mm and 95 mm, respectively.

The interference pattern of the USAF 1951 test chart on the image sensor plane was recorded in single-shot and then the object image was retrieved using the proposed technique. The recorded single-shot hologram of the transmissive USAF 1951 resolution test target is shown in Figure 5(a) which shows a nice interference modulation owning to the simple in-line configuration. The extracted four holograms at 0°, 45°, 90° and 135° four different phases are shown in Figure 5(b-e), which still own high modulated interference. By using the express of Eq. (1), complex hologram amplitude and phase were obtained as shown in Figure5(f) and 5(g), respectively. At last, a clear image of the USAF 1951 resolution chart was retrieved based on the complex hologram amplitude and phase, as shown in Fig.5(h). As a detail analysis of the red box in Fig.5(h), the intensity profiles of the line pairs show a clear and deep separation, which is shown in Fig. 5(i).

One of the most important property of digital holographic method is its capability of 3D image in single-shot exposure. To verify and explore the 3D image capability of our setup, both the USAF 1951 resolution chart and the NBS 1963 chart with 4.5 cycle/mm region at different distances are used to achieve FINCH image. The distance between two targets is about 65 mm. The results are shown in Figure 6. The same as those of Fig. 5. Fig. 6(a) is the recorded single-shot hologram of the transmissive resolution test targets. Fig.6 (b-e) are the phase-shifting holograms at 0°, 45°, 90° and 135°, respectively; Fig.6 (h-i) are the obtained final retrieved images focused on the USAF 1951 and NBS 1963, respectively. The experimental results shown in both Fig.5 and Fig.6 have proved the capability of 3D image of our novel single-shot FINCH system.

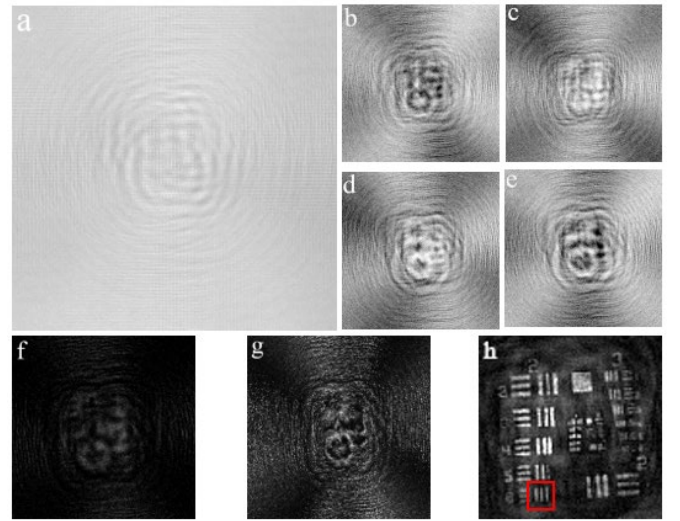

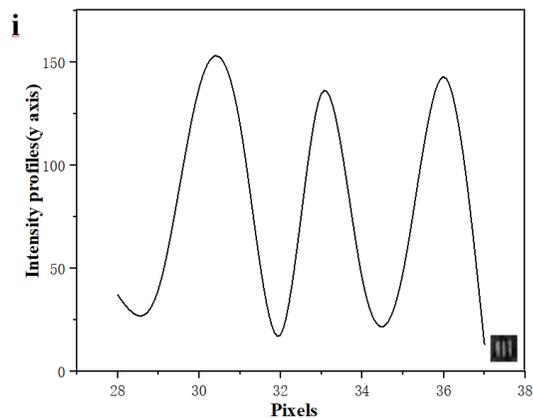

Fig.5. Recorded and Reconstructed images of a transmissive USAF 1951 test target. (a) Recorded single-shot raw hologram; (b-e) Four phase-shifting holograms; (f-g) Complex hologram amplitude and phase; (h) Reconstructed USAF1951 target image; (i)Intensity distribution profiles of the red boxes in (h).

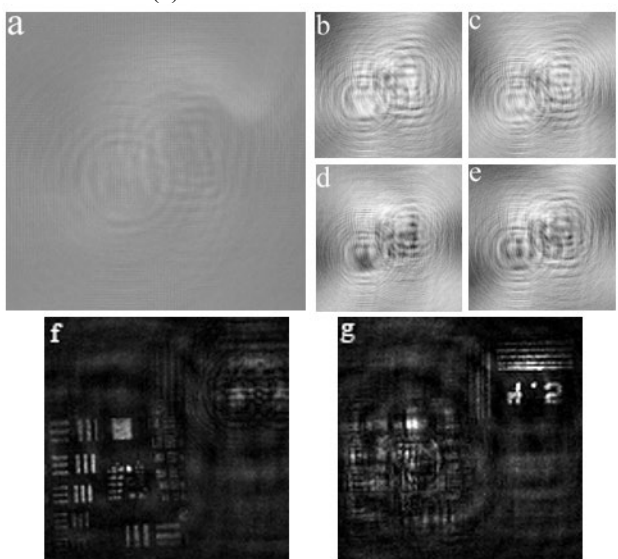

Fig.6. Recorded and Reconstructed images of two-layer volumetric test targets. (a) recorded single-shot raw hologram; (b-e) four phase-shifting holograms; (f-g) complex hologram amplitude and phase; (h) focus at the USAF 1951 target (forward); (i) focus at the NBS 1963 target (65 mm backward).

In conclusion, we have proposed a novel single-shot, in-line FINCH system by using a GP lens and a polarization camera simultaneously. The GP lens own the same ability as SLM that can separate the light into two with different wavefronts, which make the setup very simple and economical. Together with a polarization imaging camera with micro-polarizer array bonded directly to the sensor, we achieved single-shot FINCH system with compact setup. Then, the designed system is experimentally demonstrated by recording the standard resolution charts illuminated with LEDs. Experimental results showed that the bias and twin images were clearly removed by applying parallel phase-shifting to incoherent digital holography. With further optimization and customization, we believe that the combination of GP lens with the polarization camera will contribute to many research fields such as microscopy and phase metrology. More prospective applications can be expected in the near future.

**Acknowledgment.**
The authors would like to thank Prof. Guohai Situ for reading the paper. This work is supported by the National Natural Science Foundation of China (NSFC) (61521093, 61527821); Instrument Developing Project (YZ201538) and the Strategic Priority Research Program (XDB16) of the Chinese Academy of Sciences (CAS).